\documentclass[journal=jacsat,manuscript=article]{achemso}

\usepackage{chemformula} 
\usepackage[T1]{fontenc} 
\usepackage{siunitx}
\usepackage{amsmath}
\usepackage{algorithm,algorithmic}
\usepackage{graphicx}
\usepackage{caption}
\usepackage{subcaption}
\usepackage{multicol}
\usepackage{multirow}
\DeclareMathOperator*{\argmax}{argmax}
\usepackage{natbib}

\author{Mario Fordellone}
\altaffiliation{Piazzale Aldo Moro, 5 Rome (Italy)}
\email{mario.fordellone@uniroma1.it}
\affiliation[Sapienza University of Rome]
{Department of Statistical Science, Sapienza University, Rome (Italy)}
\author{Andrea Bellincontro}
\alsoaffiliation[University of Tuscia]
{Department for Innovation in Biological, Agro-food and Forest Systems, \\University of Tuscia, Viterbo (Italy)}
\author{Fabio Mencarelli}
\alsoaffiliation[University of Tuscia]
{Department for Innovation in Biological, Agro-food and Forest Systems, \\University of Tuscia, Viterbo (Italy)}

\title[]
  {Partial least squares discriminant analysis: A dimensionality reduction method to classify hyperspectral data}

\abbreviations{}

\begin{document}


\begin{abstract}
The recent development of more sophisticated spectroscopic methods allows acquisition of high dimensional datasets from which valuable information may be extracted using multivariate statistical analyses, such as dimensionality reduction and automatic classification (supervised and unsupervised). In this work, a supervised classification through a partial least squares discriminant analysis (PLS-DA) is performed on the hyperspectral data. The obtained results are compared with those obtained by the most commonly used classification approaches.
\end{abstract}

\section{Introduction}
The recent development of more sophisticated spectroscopic approaches allows the acquisition of high dimensional datasets from which valuable information may be extracted via different multivariate statistical techniques. The high data dimensionality greatly enhances the informational content of the dataset and provides an additional opportunity for the current techniques for analyzing such data \cite{jimenez1998supervised}. For example, automatic classification (clustering and/or classification) of data with similar features is an important problem in a variety of research areas such as biology, chemistry, and medicine \cite{hardy2006fine,galvan2006reversal}. When the labels of the clusters are available, a supervised classification method is applied. Several classification techniques are available and described in the literature. However, data derived by spectroscopic detection represent a hard challenge for the researcher, who faces two crucial problems: data dimensionality larger than the observations, and high correlation levels among the variables (multicollinearity).\\
\indent
Usually, in order to solve these problems \textit{(i)} a first data compression or reduction method, such as principal component analysis (PCA) is applied to shrink the number of  variables; then, a range of discriminant analysis techniques is used to solve the classification problem, while \textit{(ii)} in other cases, non-parametric classification approaches are used \cite{jimenez1998supervised,agrawal1998automatic,buhlmann2011statistics,kriegel2009clustering,ding2005classification}.\\
\indent
In this work, the dataset consists of three different varieties of olives (\textit{Moraiolo}, \textit{Dolce di Andria}, and \textit{Nocellara Etnea}) monitored during ripening up to harvest \cite{bellincontro2012feasible}. Samples contained olives from 162 trees (54 for each variety), and 601 spectral detections (i.e., dimensions/variables) were performed using a portable near infrared acousto-optically tunable filter (NIR-AOTF) device in diffuse reflectance mode from 1100 nm to 2300 nm with an interval of 2. The use of NIRS on olive fruits and related products is already known; applications for the determination of oil and
moisture content are now considered routine analyses in comparison with relatively new methodologies, such as nuclear magnetic resonance (NMR), or
more traditional analytical determinations \cite{garcia1996influence,gallardo2005application,leon2004parent,cayuela2010prediction}.\\
\indent
This paper is based on the use of partial least squares discriminant Analysis (PLS-DA). However, for comparison purposes, we also analyze the results obtained by other commonly used non-parametric classification models such as $K$-nearest neighbor (KNN), support vector machine (SVM) \cite{balabin2010gasoline,misaki2010comparison,tran2006knn,joachims2005support}, and some variants of discriminant functions for sparse data as such as diagonal linear discriminant analysis (DLDA), maximum uncertainty linear discriminant analysis (MLDA), and shrunken linear discriminant analysis (SLDA). All the three regularization techniques compute linear discriminant functions \cite{hastie1995penalized, clemmensen2011sparse,thomaz2006maximum,fisher2011improved, dudoit2002comparison,guo2006regularized}.\\
\indent
PLS-DA is a dimensionality reduction technique, a variant of partial least squares regression (PLS-R) that is used when the response variable is categorical. It is a compromise between the usual discriminant analysis and a discriminant analysis on the principal components of the predictor variables. In particular, PLS-DA instead of finding hyperplanes of maximum variance between the response and independent variables finds a linear regression model by projecting the predicted variables and the observed variables into a new space. PLS-DA can provide good insight into the causes of discrimination via weights and loadings, which gives it a unique role in exploratory data analysis, for example in metabolomics via visualization of significant variables such as metabolites or spectroscopic peaks \cite{kemsley1996discriminant,brereton2014partial,wehrens2007pls}.\\
\indent
The paper is structured as follows: in section 2 we provide a background on the most commonly used non-parametric statistical methodologies to solve the classification problem of sparse data (i.e., KNN and SVM) and an overview of different classifiers derived from linear discriminant analysis (LDA), in section 3 we focus on the PLS-DA model with a deeper examination of the PLS algorithm, in section 4 we show a comparison of the results obtained by the application of PLS-DA and those obtained by the other common classification methods, and finally in section 5 we provide some suggestions and ideas for future research.  
\section{Background}
In this section, we present a brief overview of different classifiers that have been highly successful in handling high dimensional data classification problems, starting with popular methods such as $K$-nearest neighbor (KNN) and support vector machines (SVM) \cite{dudoit2002comparison, zhang2006svm} and variants of discriminant functions for sparse data \cite{clemmensen2011sparse}. We also examine dimensionality reduction techniques and their integration with some existing algorithms (i.e., partial least squares discriminant analysis (PLS-DA)) \cite{kemsley1996discriminant,brereton2014partial}.
\subsection{$K$-nearest neighbor (KNN)}
The KNN method was first introduced by Fix and Hodges \cite{fix1989discriminatory} based on the need to perform discriminant analysis when reliable parametric estimates of probability densities are unknown or difficult to determine. In this method, a distance measure (e.g., Euclidean) is assigned between all points in the data. The data points, $K$-closest neighbors (where $K$ is the number of neighbors), are then found by analyzing a distance matrix. The $K$-closest data points are then found and analyzed in order to determine which class label is the most common among the set. Finally, the most common class label is then assigned to the data point being analyzed \cite{balabin2010gasoline}.\\
\indent
The KNN classifier is commonly based on the Euclidean distance between a test sample and the specified training samples. Formally, let $\mathbf{x}_i$ be an input sample with $J$ features ($\mathbf{x}_{i,1}, \dots, \mathbf{x}_{i,J}$), and $n$ be the total number of input samples ($i=1,  \dots, n$). The Euclidean distance between sample $\mathbf{x}_i$ and $\mathbf{x}_l$ ($l=1, \dots, n$) is defined as
\begin{equation} \label{KNN01}
d(\mathbf{x}_i, \mathbf{x}_l)=\sqrt{(\mathbf{x}_{i,1}-\mathbf{x}_{l,1})^2+\dots+(\mathbf{x}_{i,J}-\mathbf{x}_{l,J})^2}.
\end{equation}
Using the latter characteristic, the KNN classification rule is to assign to a test sample the majority category label of its $K$ nearest training samples. In other words, $K$ is usually chosen to be odd, so as to avoid ties. The $K=1$ rule is generally called the 1-nearest-neighbor classification rule.\\
\indent
Then, let $\mathbf{x}_i$ be a training sample and $\mathbf{x}_i^*$ be a test sample, and let $\omega$ be the true class of a training sample and $\hat{\omega}$ be the predicted class for a test sample ($\omega, \hat{\omega}=\, \dots, \Omega$), where $\Omega$ is the total number of classes. During the training process, only the true class $\omega$ of each training sample to train the classifier is used, while during testing the class $\hat{\omega}$ of each test sample is predicted. With 1-nearest neighbor rule, the predicted class of test sample $\mathbf{x}_i^*$ is set equal to the true class $\omega$ of its nearest neighbor, where $\mathbf{z}_i$ is a nearest neighbor to $\mathbf{x}_i^*$ if the distance 
\begin{equation} \label{KNN012}
d(\mathbf{z}_i, \mathbf{x}_i^*)=\min_j\{d(\mathbf{z}_j, \mathbf{x}_i^*)\}.
\end{equation}
For the $K$-nearest neighbors rule, the predicted class of test sample $\mathbf{x}_i^*$ is set equal to the most frequent true class among the $K$ nearest training samples.
\subsection{Support vector machine (SVM)}
The SVM approach was developed by Vapnik \cite{suykens1999least,cortes1995machine}. Synthetically, SVM is a linear method in a very high dimensional feature space that is nonlinearly related to the input space. The method maps input vectors to a higher dimensional space where a maximal separating hyperplane is constructed \cite{joachims2005support}. Two parallel hyperplanes are constructed on each side of the hyperplane that separates the data and maximizes the distance between the two parallel hyperplanes. An assumption is made that the larger the margin or distance between these parallel hyperplanes, the better the generalization error of the classifier will be.\\
\indent
SVM was initially designed for binary classification. To extend SVM to the multi-class scenario, a number of classification models were proposed \cite{wang2014multi}. Formally, given training vectors $\mathbf{x}_i\in \Re^J$, $i=1, \dots, n^*$, in two classes, and the label vector $\mathbf{Y}\in \{-1, 1\}^{n^*}$ (where $n^*$ in the size of the training samples), the support vector technique requires the solution of the following optimization problem:
\begin{equation} \label{svm01}
\begin{split}
\min_{\mathbf{w}\in H, b\in \Re, \xi_i\in \Re} &\frac{1}{2}\mathbf{w}^T\mathbf{w}+C\sum_{i=1}^{n^*}\xi_i,\\
subject \: \: to \quad &y_i(\mathbf{w}^T\varphi(\mathbf{x}_i)+b)\geq 1-\xi_i\\
\quad \quad \quad  \quad \quad \quad &\xi_i\geq 0, \quad i=1, \dots, n^*,
\end{split}
\end{equation}
where $\mathbf{w}\in \Re^J$ is the weights vector, $C\in \Re_+$ is the regularization constant, and the mapping function $\varphi$ projects the training data into a suitable feature space $H$.\\
\indent
For a $K$-class problem, many methods use a single objective function for training all $K$-binary SVMs simultaneously and maximize the margins from each class to the remaining ones \cite{wang2014multi,weston1998multi}. An example is the formulation proposed by Weston and Watkins \cite{weston1998multi}. Given a labeled training set represented by $\{(\mathbf{x}_1, y_1), \dots, (\mathbf{x}_{n^*}, y_{n^*})\}$, where $\mathbf{x}_i\in \Re^J$ and $y_i\in \{1, \dots, K\}$, this formulation is given as follows:
\begin{equation} \label{svm02}
\begin{split}
\min_{\mathbf{w}_k\in H, b\in \Re^K, \xi \in \Re^{n^*\times K}} &\frac{1}{2}\sum_{k=1}^K\mathbf{w}_k^T\mathbf{w}_k+C\sum_{i=1}^{n^*}\sum_{t\neq y_i}\xi_{i,t},\\
subject \: \: to \quad &\mathbf{w}_{y_i}^T\varphi(\mathbf{x}_i)+b_{y_i})\geq \mathbf{w}_{t}^T\varphi(\mathbf{x}_i)+b_t+2-\xi_{i,t},\\
\quad \quad \quad  \quad \quad \quad &\xi_{i,t}\geq 0, \quad i=1, \dots, n^*, \quad t\in \{1, \dots, K\}.
\end{split}
\end{equation}
The resulting decision function is given in Equation \ref{svm03} \cite{wang2014multi}.
\begin{equation} \label{svm03}
\argmax_{k} f_m(\mathbf{x})=\argmax_{k}(\mathbf{w}_{k}^T\varphi(\mathbf{x}_i)+b_{k}).
\end{equation}
\subsection{Discriminant analysis functions}
In this section we present a comprehensive overview of different classifiers derived by Linear Discriminant Analysis (LDA), and that have been highly successful in handling high dimensional data classification problems: Diagonal Linear Discriminant Analysis (DLDA), Maximum uncertainty Linear Discriminant Analysis (MLDA), and Shrunken Linear Discriminant Analysis (SLDA). All the three regularization techniques compute Linear Discriminant Functions,  by default after a preliminary variable selection step, based on alternative estimators of a within-groups covariance matrix that leads to reliable allocation rules in problems where the number of selected variables is close to, or larger than, the number of available observations.\\
\indent
The main purpose of discriminant analysis is to assign an unknown subject to one of $K$ classes on the basis of a multivariate observation $x=(x_1, \dots, x_J)^\prime$, where $J$ is the number of variables. The standard LDA procedure does not assume that the populations of the distinct groups are normally distributed, but it assumes implicitly that the true covariance matrices of each class are equal because the same within-class covariance matrix is used for all the classes considered \cite{thomaz2006maximum,wichern1992applied}. Formally, let $\mathbf{S}_b$ be the between-class covariance matrix defined as
\begin{equation} \label{da1}
\mathbf{S}_b=\sum_{k=1}^K n_k(\bar{x}_k-\bar{x})(\bar{x}_k-\bar{x})^T,
\end{equation}
and let $\mathbf{S}_w$ be the within-class covariance matrix defined as
\begin{equation} \label{da2}
\mathbf{S}_w=\sum_{k=1}^K (n_k-1)\mathbf{S}_k=\sum_{k=1}^K\sum_{i=1}^{n_k}(\bar{x}_{k,i}-\bar{x}_k)(\bar{x}_{k,i}-\bar{x}_k)^T,
\end{equation}
where $x_{k,i}$ is the $J$-dimensional pattern $i$ from the $k$-th class, $n_k$ is the number of training patterns from the $k$-th class, and $K$ is the total number of classes (or groups) considered. The vector $\bar{x}_k$ and matrix $\mathbf{S}_k$ are respectively the unbiased sample mean and sample covariance matrix of the $k$-th class, while the vector $\bar{x}$ is the overall unbiased sample mean given by
\begin{equation} \label{da3}
\bar{x}=\frac{1}{n}\sum_{k=1}^K n_k\bar{x}_k=\frac{1}{n}\sum_{k=1}^K\sum_{i=1}^{n_k}x_{k,i},
\end{equation}
where $n$ is the total number of samples $n=n_1+\dots+n_K$.\\
\indent
Then, the main objective of LDA is to find a projection matrix (here defined as $\mathbf{P}_{LDA}$) that maximizes the ratio of the determinant of the between-class scatter matrix to the determinant of the within-class scatter matrix (Fisher's  criterion). Formally, 
\begin{equation} \label{da4}
\mathbf{P}_{LDA}=\argmax_{\mathbf{P}} \frac{det\left(\mathbf{P}^T\mathbf{S}_b\mathbf{P}\right)}{det\left(\mathbf{P}^T\mathbf{S}_w\mathbf{P}\right)}.
\end{equation}
It has been shown \cite{devijver1982pattern} that Equation (\ref{da4}) is in fact the solution of the following eigenvector system problem:
\begin{equation} \label{da5}
\mathbf{S}_b\mathbf{P}-\mathbf{S}_w\mathbf{P}\Lambda=0.
\end{equation}
Note that by multiplying both sides by $\mathbf{S}_w^{-1}$, Equation (\ref{da5}) can be rewritten as
\begin{equation} \label{da6}
\begin{split}
&\mathbf{S}_w^{-1}\mathbf{S}_b\mathbf{P}-\mathbf{S}_w^{-1}\mathbf{S}_w\mathbf{P}\Lambda=0\\
&\mathbf{S}_w^{-1}\mathbf{S}_b\mathbf{P}-\mathbf{P}\Lambda=0\\
&(\mathbf{S}_w^{-1}\mathbf{S}_b)\mathbf{P}=\mathbf{P}\Lambda,
\end{split}
\end{equation}
where $\mathbf{P}$ and $\Lambda$ are respectively the eigenvector and eigenvalue matrices of the $\mathbf{S}_w^{-1}\mathbf{S}_b$ matrix. These eigenvectors are primarily used for dimensionality reduction, as in principal component analysis (PCA) \cite{rao1948utilization}.\\
\indent
However, the performance of the standard LDA can be seriously degraded if there are only a limited number of total training observations $n$ compared to the number of dimensions of the feature space $J$. In this context, in fact the $\mathbf{S}_w$ matrix becomes singular. To solve this problem, Yu and Yang \cite{thomaz2006maximum, yu2001direct} have developed a direct LDA algorithm (called DLDA) for high dimensional data with application to face recognition that diagonalizes simultaneously the two symmetric matrices $\mathbf{S}_w$ and $\mathbf{S}_b$. The idea of DLDA is to discard the null space of $\mathbf{S}_b$ by diagonalizing $\mathbf{S}_b$ first and then diagonalizing $\mathbf{S}_w$.\\
\indent
The following steps describe the DLDA algorithm for calculating the projection matrix $\mathbf{P}_{DLDA}$:\\[2ex]
\textbf{1}. diagonalize $\mathbf{S}_b$, that is, calculate the eigenvector matrix $\mathbf{V}$ such that $\mathbf{V}^T\mathbf{S}_b\mathbf{V}=\Lambda$;\\
\textbf{2}. let $\mathbf{Y}$ be a sub-matrix with the first $m$ columns of $\mathbf{V}$ corresponding to the $\mathbf{S}_b$ largest eigenvalues, where $m\leq rank(\mathbf{S}_b)$. Calculate the diagonal $m\times m$ sub-matrix of the eigenvalues of $\Lambda$ as $\mathbf{D}_b=\mathbf{Y}^T\mathbf{S}_b\mathbf{Y}$;\\ 
\textbf{3}. let $\mathbf{Z}=\mathbf{YD}_b^{-1/2}$ be a whitening transformation of $\mathbf{S}_b$ that reduces its dimensionality from $J$ to $m$ (where $\mathbf{Z}^T\mathbf{S}_b\mathbf{Z}=\mathbf{I}$). Diagonalize $\mathbf{Z}^T\mathbf{S}_w\mathbf{Z}$, that is, compute $\mathbf{U}$ and $\mathbf{D}_w$ such that $\mathbf{U}^T(\mathbf{Z}^T\mathbf{S}_w\mathbf{Z})\mathbf{U}=\mathbf{D}_w$;\\
\textbf{4}. calculate the projection matrix as $\mathbf{P}_{DLDA}=\mathbf{D}_w^{-1/2}\mathbf{U}^T\mathbf{Z}^T$.\\[2ex]
Note that by replacing the between-class covariance matrix $\mathbf{S}_b$ with total covariance matrix $\mathbf{S}_T$ ($\mathbf{S}_T=\mathbf{S}_b+\mathbf{S}_w$), the first two steps of the algorithm become exactly the PCA dimensionality reduction technique \cite{yu2001direct}.\\
\indent
Two other approaches commonly used to avoid both the critical singularity and instability issues of the within-class covariance matrix $\mathbf{S}_w$ are SLDA and the MLDA \cite{thomaz2006maximum}. Firstly, it is important to note that the within-class covariance matrix $\mathbf{S}_w$ is essentially the standard pooled covariance matrix $\mathbf{S}_p$ multiplied by the scalar $(n-K)$. Then,
\begin{equation} \label{da7}
\mathbf{S}_w=\sum_{k=1}^K (n_k-1)\mathbf{S}_k=(n-K)\mathbf{S}_p.
\end{equation}
From this property, the key idea of some regularization proposals of LDA \cite{guo2006regularized, campbell1980shrunken, peck1982use} is to replace the pooled covariance matrix $\mathbf{S}_p$ of the within-class covariance matrix $\mathbf{S}_w$ with the following convex combination:
\begin{equation} \label{da8}
\hat{\mathbf{S}}_p(\gamma)=(1-\gamma)\mathbf{S}_p+\gamma \bar{\lambda}\mathbf{I},
\end{equation}
where $\gamma \in [0, 1]$ is the shrinkage parameter, which can be selected to maximize the leave-one-out classification accuracy \cite{cawley2003efficient}, $\mathbf{I}$ is the identity matrix, and $\bar{\lambda}=J^{-1}\sum_{j=1}^J \lambda_j$ is the average eigenvalue, which can be written as $J^{-1} trace (\mathbf{S}_p)$. This regularization approach, called SLDA, would have the effect of decreasing the larger eigenvalues and increasing the smaller ones, thereby counteracting the biasing inherent in eigenvalue sample-based estimation \cite{thomaz2006maximum,hastie1995penalized}.\\
\indent
In contrast, in the MLDA method a multiple of the identity matrix determined by selecting the largest dispersions regarding the $\mathbf{S}_p$ average eigenvalue is used. In particular, if we replace the pooled covariance matrix $\mathbf{S}_p$ of the covariance matrix $\mathbf{S}_w$ (shown in Equation (\ref{da7})) with a covariance estimate of the form $\hat{\mathbf{S}}_p(\delta)=\mathbf{S}_p+\delta\mathbf{I}$ (where $\delta \geq 0$ is an identity matrix multiplier), then the eigen-decomposition of a combination of the covariance matrix $\mathbf{S}_p$ and the $J\times J$ identity matrix $\mathbf{I}$ can be written as 
\begin{equation} \label{da9}
\begin{split}
\hat{\mathbf{S}}_p(\delta)&=\mathbf{S}_p+\delta\mathbf{I}\\
&=\sum_{j=1}^r \lambda_j\phi_j(\phi_j)^T+\delta \sum_{j=1}^J\phi_j(\phi_j)^T\\
&=\sum_{j=1}^r (\lambda_j+\delta)\phi_j(\phi_j)^T+\sum_{j=1}^J\delta \phi_j(\phi_j)^T,
\end{split}
\end{equation}
where $r$ is the rank of $\mathbf{S}_p$ (note that $r\leq J$), $\lambda_j$ is the $j$-th eigenvalue of $\mathbf{S}_p$, $\phi_j$ is the $j$-th corresponding eigenvector, and $\delta$ is the identity matrix multiplier previously defined. In fact, in Equation (\ref{da9}) the identity matrix is defined as $\mathbf{I}=\sum_{j=1}^J\phi_j(\phi_j)^T$. Now, given the convex combination shown in Equation (\ref{da8}), the eigen-decomposition can be written as     
\begin{equation} \label{da10}
\begin{split}
\hat{\mathbf{S}}_p(\gamma)&=(1-\gamma)\mathbf{S}_p+\gamma \bar{\lambda}\mathbf{I}\\
&=(1-\gamma)\sum_{j=1}^r \lambda_j\phi_j(\phi_j)^T+\gamma \sum_{j=1}^J\bar{\lambda}\phi_j(\phi_j)^T.
\end{split}
\end{equation}
The steps of the MLDA algorithm are shown follows:\\[2ex]
\textbf{1}. Find the $\Phi$ eigenvectors matrix and $\Lambda$ eigenvalues matrix ff $\mathbf{S}_p$, where $\mathbf{S}_p=(n-K)\mathbf{S}_w$ (from Equation (\ref{da7}));\\
\textbf{2}. Calculate $\mathbf{S}_p$ average eigenvalues as $J^{-1} trace (\mathbf{S}_p)$;\\
\textbf{3}. Construct a new matrix of eigenvalues based on the following largest dispersion values :
$$
\Lambda^{*}=diag\left[max(\lambda_1, \bar{\lambda}), \dots, max(\lambda_J,\bar{\lambda})\right];
$$
\textbf{4}. Define the revised within-class covariance matrix:\\
$$
\mathbf{S}_w^*=(n-K)\mathbf{S}_p^*=(n-K)(\Phi \Lambda^*\Phi^T).
$$
Then, the MLDA approach is based on replacing $\mathbf{S}_w$ with $\mathbf{S}_w^*$ in the Fisher’s criterion formula described in Equation (\ref{da4}).
\section{Partial Least Squares Discriminant Analysis (PLS-DA)}
Multivariate regression methods like principal component regression (PCR) and partial least squares regression (PLS-R) enjoy large popularity in a wide range of fields and are mostly used in situations where there are many, possibly correlated, predictor variables and relatively few samples, a situation that is common, especially in chemistry, where developments in spectroscopy since the seventies have revolutionized chemical analysis \cite{wehrens2007pls,perez2003prediction}. In fact, the origin of PLSR lies in chemistry \cite{wehrens2007pls, martens2001reliable,wold2001personal}.\\
\indent
In practice, there are not many differences between the use of PCR and PLS-R; in most situations, the methods achieve similar prediction accuracies. Note that with the same number of latent variables, PLS-R will cover more of the variation in $\mathbf{Y}$ and PCR will cover more of the variation in $\mathbf{X}$. \cite{wehrens2007pls}.\\
\indent
Partial least squares discriminant Analysis (PLS-DA) is a variant of PLS-R that can be used when the response variable $\mathbf{Y}$ is categorical. Under certain circumstances, PLS-DA provides the same results as the classical approach of Euclidean distance to centroids (EDC) \cite{davies1979cluster} and under other circumstances, the same as that of linear discriminant analysis (LDA) \cite{izenman2013linear}. However, in different contexts this technique is specially suited to deal with models with many more predictors than observations and with multicollinearity, two of the main problems encountered when analyzing hyperspectral detection data \cite{perez2003prediction}.
\subsection{Model and algorithm}
PLS-DA is derived from PLS-R, where the response vector $\mathbf{Y}$ assumes discrete values. In the usual multiple linear regression model (MLR) approach we have
\begin{equation} \label{PLS01}
\mathbf{Y}=\mathbf{XB}+\mathbf{F},
\end{equation}
where $\mathbf{X}$ is the $n\times J$ data matrix, $\mathbf{B}$ is the $J\times 1$ regression coefficients matrix, $\mathbf{F}$ is the $n\times 1$ error vector, and $\mathbf{Y}$ is the $n\times 1$ response variable vector. In this approach, the least squares solution is given by $\mathbf{B}=(\mathbf{X^TX})^{-1}\mathbf{X^TY}$.\\
\indent
In many cases, the problem is the singularity of the $\mathbf{X^TX}$ matrix (e.g., when there are multicollinearity problems in the data or the number of predictors is larger than the number of observations). Both PLS-R and PLS-DA solve this problem by decomposing the data matrix $\mathbf{X}$ into $P$ orthogonal scores $\mathbf{T}$ ($n\times P$) and loadings matrix $\mathbf{P}$ ($J\times P$), and the response vector $\mathbf{Y}$ into $P$ orthogonal scores $\mathbf{T}$ ($n\times P$) and loadings matrix $\mathbf{Q}$ ($1\times P$). Then, let $\mathbf{E}$ and $\mathbf{F}$ be the $n\times J$ and $n\times 1$ error matrices associated with the data matrix $\mathbf{X}$ and response vector $Y$, respectively. There are two fundamental equations in the PLS-DA model:
\begin{equation} \label{PLS02}
\begin{split}
&\mathbf{X}=\mathbf{TP^T}+\mathbf{E}\\
&\mathbf{Y}=\mathbf{TQ^T}+\mathbf{F}.
\end{split}
\end{equation}
Now, if we define a $J\times P$ weights matrix $\mathbf{W}$, we can write the scores matrix as 
\begin{equation} \label{PLS03}
\mathbf{T}=\mathbf{XW}(\mathbf{P^TW})^{-1}, 
\end{equation}
and by substituting it into the PLS-DA model, we obtain
\begin{equation} \label{PLS04}
\mathbf{Y}=\mathbf{XW}(\mathbf{P^TW})^{-1}\mathbf{Q^T}+\mathbf{F},
\end{equation}
where the regression coefficient vector $\mathbf{B}$ is given by
\begin{equation} \label{PLS05}
\hat{\mathbf{B}}=\mathbf{W}(\mathbf{P^TW})^{-1}\mathbf{Q^T}.
\end{equation}
In this way, an unknown sample value of $\mathbf{Y}$ can be predicted by $\hat{\mathbf{Y}}=\mathbf{X}\hat{\mathbf{B}}$, i.e. $\hat{\mathbf{Y}}=\mathbf{X}\mathbf{W}(\mathbf{P^TW})^{-1}\mathbf{Q^T}$. The PLS-DA algorithm estimates the matrices $\mathbf{W}$, $\mathbf{T}$, $\mathbf{P}$, and $\mathbf{Q}$ through the following steps \cite{brereton2014partial}.
\begin{algorithm}[H]
  \algsetup{linenosize=\small}
  \small
\begin{algorithmic}[1]
\STATE Fixed $P$, initialize the residuals matrices $\mathbf{E_0}=\mathbf{X}$ and $\mathbf{F_0}=\mathbf{Y}$;
\FOR{$p=1$ to $P$}
\STATE Calculate PLS weights vector\\ $\mathbf{W_{p}}=\mathbf{E_{0}^T}\mathbf{F_{0}}$;
\STATE Calculate and normalize scores vector\\ $\mathbf{T_p}=\mathbf{E_0}\mathbf{W_p}(\mathbf{W_p^T E_0^T E_0 W_p})^{-1/2}$ ;
\STATE Calculate the $\mathbf{X}$ loadings vector\\ $\mathbf{P_p}=\mathbf{E_0^T T_p}$;
\STATE Calculate $\mathbf{Y}$ loading\\ $\mathbf{Q_p}=\mathbf{F_0^T T_p}$;
\STATE Update the $\mathbf{X}$ residuals vector\\ $\mathbf{E_0}=\mathbf{E_{0}}-\mathbf{T_p P_p^T}$;
\STATE Update the $\mathbf{Y}$ residuals vector\\ $\mathbf{F_0}=\mathbf{F_{0}}-\mathbf{T_p Q_p^T}$;
\ENDFOR
\STATE Obtain output matrices $\mathbf{W}$, $\mathbf{T}$, $\mathbf{P}$, $\mathbf{Q}$. 
\end{algorithmic}
\caption{Partial Least Squares}
\end{algorithm}
\section{Application to real data}
In this section we show an application of the method to real data. In particular, we compare the results obtained by partial least squares discriminant analysis (PLS-DA) and the other classification techniques discussed in Section 2.
\subsection{Dataset} 
The dataset consists of 162 drupes of olives harvested in 2010 belonging to three different cultivars (response variable): 54 \textit{Dolce di Andria} (low phenolic concentration), 54 \textit{Moraiolo} (high phenolic concentration), and 54 \textit{Nocellara Etnea} (medium phenolic concentration). Spectral detection is performed using a portable NIR device (diffuse reflectance mode) in the 1100--2300 nm wavelength range, with 2 nm wavelength increments (601 observed variables) \cite{bellincontro2012feasible}.
{\begin{figure}[!h]
\centering
\includegraphics[scale=0.5]{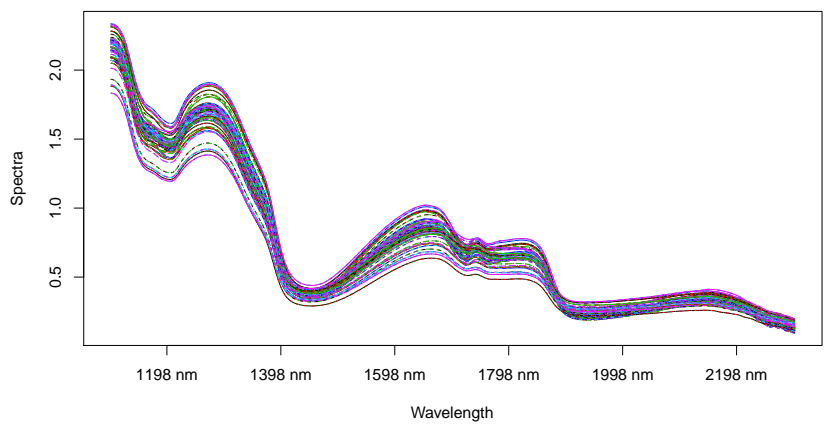}
\caption{Representation of spectral detections performed on the 1100--2300 nm wavelength range}
\label{fig:01}
\end{figure}
\subsection{Principal results} 
In order to evaluate the prediction capability of the model, the entire data set has been randomly divided into a \textit{training set} composed of 111 balanced observations (i.e., about 70\% of the entire sample, with each class composed of 37 elements), and a \textit{test set} (drawn from the sample) composed of 51 observations balanced across the three cultivars (i.e., about 30\% of the entire sample and each class composed by 17 elements) \cite{guyon1998size}.\\
\indent
The first step of the analysis consists in selecting the optimal number of components $P$, i.e., the number of latent scores to consider for representing the original variable space. For this purpose, the latent subspace must explain the largest possible proportion of the total variance to guarantee the best model estimation. Table \ref{tab:01} shows the proportion of the total variance explained by the first five components identified by PLS-DA.  
\begin{table}[h!]
  \begin{center}
    \caption{Cumulative proportion of the total variance explained by the first five components (percent values)}
    \label{tab:01}
    \begin{tabular}{l c c c c c}
     &	Comp. 1	&	Comp. 2	&	Comp. 3	&	Comp. 4	&	Comp. 5\\
	\hline
Exp.Variance	&	61.152	&	35.589	&	0.892	&	0.982	&	1.167\\
Cum. Sum		&	61.152	&	96.741	&	97.633	&	98.615	&	99.782\\
	\hline
	\end{tabular}
  	\end{center}
\end{table}
\\
\indent
The table shows that the first two components explain about 97\% of the total variance, and only the first two latent scores have a significant contribution. Thus, it seems that the best latent subspace is represented by the plane composed of the first two identified components. However, in order to guarantee the best model estimate, it is also useful to understand its prediction quality with regard to the different subspace dimensions. In other words, the selection of the optimal number of components must be related to some criterion that ensures the maximum prediction quality of the estimated model. In this paper, we propose the maximization of the chi-squared test applied on the comparison between the real training partition and the predicted training partition \cite{rao1981analysis}. Figure \ref{fig:02} represents the chi-squared values for different numbers of components (i.e., from 2 to 10 selected components). 
{\begin{figure}[!h]
\centering
\includegraphics[scale=0.5]{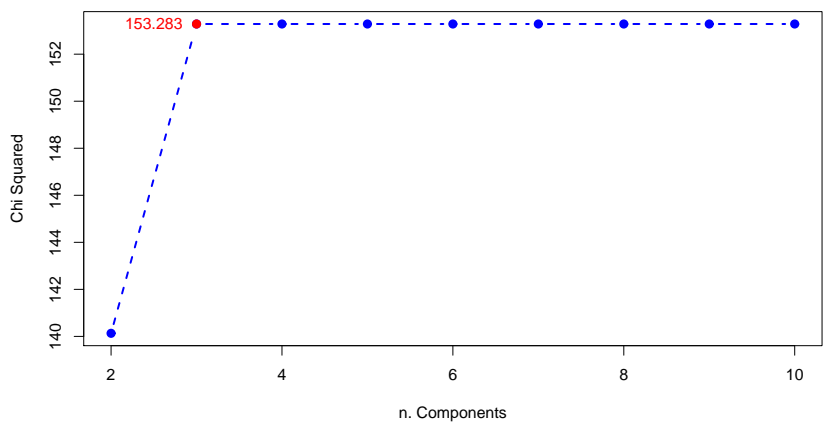}
\caption{Chi-squared values with respect to different choices of components number}
\label{fig:02}
\end{figure}
\\
\indent
In the scree-plot shown in Figure \ref{fig:02}, the chi-squared criterion suggests $P=3$ as the optimal number of components, where the maximum value of the chi-squared test is equal to 153.28. Then, we can select three components to estimate the model, but we can use the plane composed of the first two latent scores to represent the estimated groups (i.e., using 97\% of the total information in the data).\\
\indent
Figure \ref{fig:03} shows the loadings distributions and the squared of the loadings distributions of the three $\mathbf{X}$s' latent scores, measured on all the observed variables (i.e., on the 1100--2300 nm wavelength range).
{\begin{figure}[!h]
\centering
\includegraphics[scale=0.5]{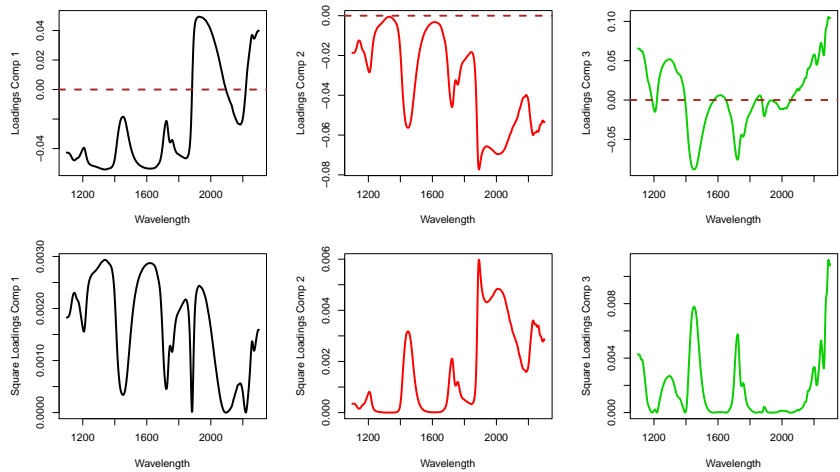}
\caption{The loadings distributions (top) and squared loadings distributions (bottom) of the three latent scores measured on all the observed variables}
\label{fig:03}
\end{figure}
\\
\indent
By observing the behavior of the loadings, we can say that the wavelengths from about 1100 nm to about 1500 nm have a high negative contribution to the first two components, while they have a positive contribution to the third component; the wavelengths from about 1500 nm to about 1900 nm have a negative contribution to all three components, with the largest contribution to the first component; finally, the wavelengths from about 1900 nm to about 2300 nm have a positive contribution to both the first and the third component, while they have a negative contribution to the second component.
\\
\indent
Now, we compare the classification results obtained by the PLS-DA procedure with results obtained by other classifiers, including $K$-nearest neighbor (KNN, $R$ package:\cite{knn_package}), support vector machine (SVM, $R$ package \cite{svm_package}), diagonal linear discriminant analysis (DLDA, $R$ package \cite{LDA_sparse_package}), maximum uncertainty linear discriminant analysis (MLDA, $R$ package \cite{LDA_sparse_package}), and shrunken linear discriminant analysis (SLDA, $R$ package \cite{LDA_sparse_package}). For the measurement of the model prediction quality, we have used \textit{mis classification rate} (MIS), \textit{adjusted Rand Index} (ARI) \cite{hubert1985comparing}, and the \textit{chi-squared} test ($\chi^2$). The three measures have been computed on the comparison between the real data partition and the predicted partition.\\
\indent
Formally, let Table \ref{tab:02} (here called $T$) be the $K\times K$ confusion matrix where the real data partition and the predicted partition have been compared, $MIS=1-n^{-1}\left[\sum_{r=1}^{R}\sum_{=1}^{C} n_{rc}\right]$, while $ARI=\frac{\sum_{r=1}^{R}\sum_{=1}^{C}\binom{n_{rc}}{2}-\binom{n}{2}^{-1}\sum_{r=1}^{R}\binom{n_{r.}}{2} \sum_{c=1}^{C}\binom{n_{.c}}{2}}{\frac{1}{2}\left[ \sum_{r=1}^{R}\binom{n_{r.}}{2}+ \sum_{c=1}^{C}\binom{n_{.c}}{2}\right ]-\binom{n}{2}^{-1}\sum_{r=1}^{R}\binom{n_{r.}}{2} \sum_{c=1}^{C}\binom{n_{.c}}{2}}$. 
\begin{table}[h!]
  \begin{center}
    \caption{An example of a confusion matrix between the real data partition and the predicted partition}
    \label{tab:02}
\begin{tabular}{c c | c c c c | c }
        &      \multicolumn{5}{ c }{\hspace{24pt}  Predicted partition} &  \\
 & & $P_1$  & \multicolumn{2}{c}{$\cdots$}   & $P_C$ & \\  
\cline{2-7} 
\multirow{1}{*}{Real partition}
                & $R_1$     &   $n_{11}$    & \multicolumn{2}{c}{$\cdots$}   & $n_{1C}$ & $n_{1\cdot}$ \\ 
                & $\vdots$  &   $\vdots$     & \multicolumn{2}{c}{$\ddots$}   & $\vdots$  &  $\vdots$ \\ 
                & $R_R$     &   $n_{R1}$  & \multicolumn{2}{c}{$\cdots$}   & $n_{RC}$ & $n_{R\cdot}$ \\ 
\cline{2-7}
               &    &   $n_{\cdot 1}$        & \multicolumn{2}{c}{$\cdots$}   & $n_{\cdot C}$ & $n$ \\ 
\end{tabular} 
  	\end{center}
\end{table}
\\
\indent
Table \ref{tab:03} shows the results for the quality of the model predictions obtained on the training set and the test set.
\begin{table}[h!]
  \begin{center}
    \caption{Model prediction quality computed on the training set and the test set}
    \label{tab:03}
    \begin{tabular}{l | c c c || c c c |}
    & \multicolumn{3}{c||}{\textit{Training set}} & \multicolumn{3}{c|}{\textit{Test set}}\\
	&	MIS	&	ARI	&	$\chi^2$	&	MIS	&	ARI	&	$\chi^2$\\
    \hline
PLS-DA	&	0.002	&	0.880	&	153.283	&	0.008	&	0.710	&	77.182\\
KNN	&	0.027	&	0.755	&	151.744	&	0.157	&	0.625	&	65.294\\
SVM	&	0.072	&	0.797	&	152.688	&	0.137	&	0.615	&	69.750\\
DLDA	&	0.241	&	0.368	&	101.599	&	0.255	&	0.351	&	46.714\\
MLDA	&	0.078	&	0.734	&	149.577	&	0.010	&	0.699	&	72.311\\
SLDA	&	0.005	&	0.712	&	150.456	&	0.011	&	0.702	&	75.899\\
	\hline
	\end{tabular}
  	\end{center}
\end{table}
\\
\indent
From the results, we can see that PLS-DA has the best performance on both the training set and the test set. This result is confirmed by the representation of the predicted partition on the first two $\mathbf{X}$s' latent scores (i.e., on about 97\% of the total data variance) as shown in Figures \ref{fig:04} and \ref{fig:05} (training set and the test set, respectively). In fact, we can see that, with respect to the other studied methodologies, PLS-DA identifies more homogeneous and better-separated classes. 
{\begin{figure}[!h]
\centering
\includegraphics[scale=0.6]{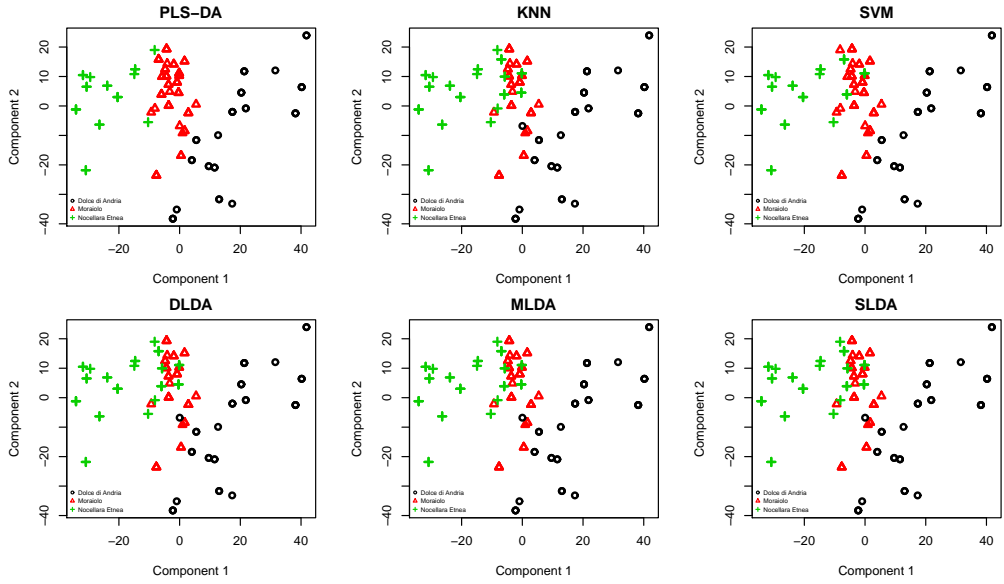}
\caption{Representation of the predicted partition on the first two latent scores (training set)}
\label{fig:04}
\end{figure}
{\begin{figure}[!h]
\centering
\includegraphics[scale=0.6]{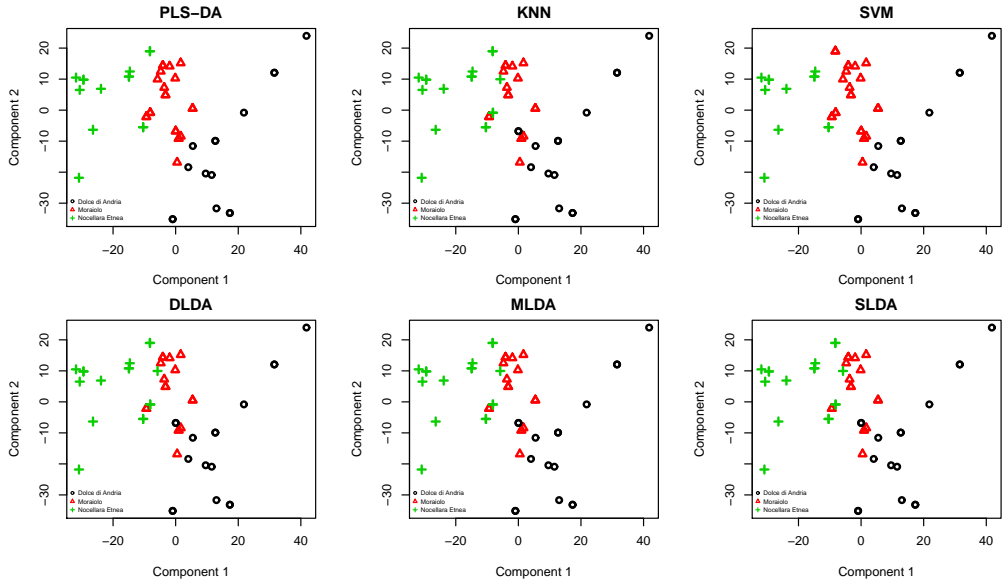}
\caption{Representation of the predicted partition on the first two latent scores (test set)}
\label{fig:05}
\end{figure}
\section{Concluding remarks}
Data acquired via spectroscopic detection represent a hard challenge for researchers, who face two crucial problems: data dimensionality larger than the number of observations, and high correlation levels among the variables. In this paper, partial least squares discriminant analysis (PLS-DA) modeling was proposed as a method to classify hyperspectral data. The results obtained on real data show that PLS-DA identifies classes that are more homogeneous and better-separated than other commonly used methods, such as non-parametric classifiers and other discriminant functions.\\
\indent
Moreover, we think that PLS-DA is a very important tool in terms of dimensionality reduction, as it can maximize the total variance of data using just a few components (i.e., the $\mathbf{X}$s' latent scores). In fact, the PLS-DA components enable a good graphical representation of the partition, which is not possible with other approaches.\\
\indent
In future studies, the use of PLS for unsupervised classification could be a useful tool when both the number and structure of the groups are unknown. 
\bibliography{sample}

\providecommand{\latin}[1]{#1}
\providecommand*\mcitethebibliography{\thebibliography}
\csname @ifundefined\endcsname{endmcitethebibliography}
  {\let\endmcitethebibliography\endthebibliography}{}
\begin{mcitethebibliography}{50}
\providecommand*\natexlab[1]{#1}
\providecommand*\mciteSetBstSublistMode[1]{}
\providecommand*\mciteSetBstMaxWidthForm[2]{}
\providecommand*\mciteBstWouldAddEndPuncttrue
  {\def\EndOfBibitem{\unskip.}}
\providecommand*\mciteBstWouldAddEndPunctfalse
  {\let\EndOfBibitem\relax}
\providecommand*\mciteSetBstMidEndSepPunct[3]{}
\providecommand*\mciteSetBstSublistLabelBeginEnd[3]{}
\providecommand*\EndOfBibitem{}
\mciteSetBstSublistMode{f}
\mciteSetBstMaxWidthForm{subitem}{(\alph{mcitesubitemcount})}
\mciteSetBstSublistLabelBeginEnd
  {\mcitemaxwidthsubitemform\space}
  {\relax}
  {\relax}

\bibitem[Jimenez and Landgrebe(1998)Jimenez, and
  Landgrebe]{jimenez1998supervised}
Jimenez,~L.~O.; Landgrebe,~D.~A. Supervised classification in high-dimensional
  space: geometrical, statistical, and asymptotical properties of multivariate
  data. \emph{IEEE Transactions on Systems, Man, and Cybernetics, Part C
  (Applications and Reviews)} \textbf{1998}, \emph{28}, 39--54\relax
\mciteBstWouldAddEndPuncttrue
\mciteSetBstMidEndSepPunct{\mcitedefaultmidpunct}
{\mcitedefaultendpunct}{\mcitedefaultseppunct}\relax
\EndOfBibitem
\bibitem[Hardy \latin{et~al.}(2006)Hardy, Maggia, Bandou, Breyne, Caron,
  CHEVALLIER, Doligez, Dutech, Kremer, LATOUCHE-HALL{\'E}, \latin{et~al.}
  others]{hardy2006fine}
others,, \latin{et~al.}  Fine-scale genetic structure and gene dispersal
  inferences in 10 Neotropical tree species. \emph{Molecular ecology}
  \textbf{2006}, \emph{15}, 559--571\relax
\mciteBstWouldAddEndPuncttrue
\mciteSetBstMidEndSepPunct{\mcitedefaultmidpunct}
{\mcitedefaultendpunct}{\mcitedefaultseppunct}\relax
\EndOfBibitem
\bibitem[Galvan \latin{et~al.}(2006)Galvan, Gorostiza, Banwait, Ataie,
  Logvinova, Sitaraman, Carlson, Sagi, Chevallier, Jin, \latin{et~al.}
  others]{galvan2006reversal}
others,, \latin{et~al.}  Reversal of Alzheimer's-like pathology and behavior in
  human APP transgenic mice by mutation of Asp664. \emph{Proceedings of the
  National Academy of Sciences} \textbf{2006}, \emph{103}, 7130--7135\relax
\mciteBstWouldAddEndPuncttrue
\mciteSetBstMidEndSepPunct{\mcitedefaultmidpunct}
{\mcitedefaultendpunct}{\mcitedefaultseppunct}\relax
\EndOfBibitem
\bibitem[Agrawal \latin{et~al.}(1998)Agrawal, Gehrke, Gunopulos, and
  Raghavan]{agrawal1998automatic}
Agrawal,~R.; Gehrke,~J.; Gunopulos,~D.; Raghavan,~P. \emph{Automatic subspace
  clustering of high dimensional data for data mining applications}; ACM, 1998;
  Vol.~27\relax
\mciteBstWouldAddEndPuncttrue
\mciteSetBstMidEndSepPunct{\mcitedefaultmidpunct}
{\mcitedefaultendpunct}{\mcitedefaultseppunct}\relax
\EndOfBibitem
\bibitem[B{\"u}hlmann and Van De~Geer(2011)B{\"u}hlmann, and Van
  De~Geer]{buhlmann2011statistics}
B{\"u}hlmann,~P.; Van De~Geer,~S. \emph{Statistics for high-dimensional data:
  methods, theory and applications}; Springer Science \& Business Media,
  2011\relax
\mciteBstWouldAddEndPuncttrue
\mciteSetBstMidEndSepPunct{\mcitedefaultmidpunct}
{\mcitedefaultendpunct}{\mcitedefaultseppunct}\relax
\EndOfBibitem
\bibitem[Kriegel \latin{et~al.}(2009)Kriegel, Kr{\"o}ger, and
  Zimek]{kriegel2009clustering}
Kriegel,~H.-P.; Kr{\"o}ger,~P.; Zimek,~A. Clustering high-dimensional data: A
  survey on subspace clustering, pattern-based clustering, and correlation
  clustering. \emph{ACM Transactions on Knowledge Discovery from Data (TKDD)}
  \textbf{2009}, \emph{3}, 1\relax
\mciteBstWouldAddEndPuncttrue
\mciteSetBstMidEndSepPunct{\mcitedefaultmidpunct}
{\mcitedefaultendpunct}{\mcitedefaultseppunct}\relax
\EndOfBibitem
\bibitem[Ding and Gentleman(2005)Ding, and Gentleman]{ding2005classification}
Ding,~B.; Gentleman,~R. Classification using generalized partial least squares.
  \emph{Journal of Computational and Graphical Statistics} \textbf{2005},
  \emph{14}, 280--298\relax
\mciteBstWouldAddEndPuncttrue
\mciteSetBstMidEndSepPunct{\mcitedefaultmidpunct}
{\mcitedefaultendpunct}{\mcitedefaultseppunct}\relax
\EndOfBibitem
\bibitem[Bellincontro \latin{et~al.}(2012)Bellincontro, Taticchi, Servili,
  Esposto, Farinelli, and Mencarelli]{bellincontro2012feasible}
Bellincontro,~A.; Taticchi,~A.; Servili,~M.; Esposto,~S.; Farinelli,~D.;
  Mencarelli,~F. Feasible application of a portable NIR-AOTF tool for on-field
  prediction of phenolic compounds during the ripening of olives for oil
  production. \emph{Journal of agricultural and food chemistry} \textbf{2012},
  \emph{60}, 2665--2673\relax
\mciteBstWouldAddEndPuncttrue
\mciteSetBstMidEndSepPunct{\mcitedefaultmidpunct}
{\mcitedefaultendpunct}{\mcitedefaultseppunct}\relax
\EndOfBibitem
\bibitem[Garcia \latin{et~al.}(1996)Garcia, Seller, and
  Perez-Camino]{garcia1996influence}
Garcia,~J.~M.; Seller,~S.; Perez-Camino,~M.~C. Influence of fruit ripening on
  olive oil quality. \emph{Journal of agricultural and food chemistry}
  \textbf{1996}, \emph{44}, 3516--3520\relax
\mciteBstWouldAddEndPuncttrue
\mciteSetBstMidEndSepPunct{\mcitedefaultmidpunct}
{\mcitedefaultendpunct}{\mcitedefaultseppunct}\relax
\EndOfBibitem
\bibitem[Gallardo \latin{et~al.}(2005)Gallardo, Osorio, and
  Sanchez]{gallardo2005application}
Gallardo,~L.; Osorio,~E.; Sanchez,~J. Application of near infrared spectroscopy
  (NIRS) for the real-time determination of moisture and fat contents in olive
  pastes and wastes of oil extraction. \emph{Alimentaci{\'o}n Equipos y
  Tecnologia} \textbf{2005}, \emph{24}, 85--89\relax
\mciteBstWouldAddEndPuncttrue
\mciteSetBstMidEndSepPunct{\mcitedefaultmidpunct}
{\mcitedefaultendpunct}{\mcitedefaultseppunct}\relax
\EndOfBibitem
\bibitem[Le{\'o}n \latin{et~al.}(2004)Le{\'o}n, Garrido-Varo, and
  Downey]{leon2004parent}
Le{\'o}n,~L.; Garrido-Varo,~A.; Downey,~G. Parent and harvest year effects on
  near-infrared reflectance spectroscopic analysis of olive (Olea europaea L.)
  fruit traits. \emph{Journal of agricultural and food chemistry}
  \textbf{2004}, \emph{52}, 4957--4962\relax
\mciteBstWouldAddEndPuncttrue
\mciteSetBstMidEndSepPunct{\mcitedefaultmidpunct}
{\mcitedefaultendpunct}{\mcitedefaultseppunct}\relax
\EndOfBibitem
\bibitem[Cayuela and Camino(2010)Cayuela, and Camino]{cayuela2010prediction}
Cayuela,~J.~A.; Camino,~M. d. C.~P. Prediction of quality of intact olives by
  near infrared spectroscopy. \emph{European journal of lipid science and
  technology} \textbf{2010}, \emph{112}, 1209--1217\relax
\mciteBstWouldAddEndPuncttrue
\mciteSetBstMidEndSepPunct{\mcitedefaultmidpunct}
{\mcitedefaultendpunct}{\mcitedefaultseppunct}\relax
\EndOfBibitem
\bibitem[Balabin \latin{et~al.}(2010)Balabin, Safieva, and
  Lomakina]{balabin2010gasoline}
Balabin,~R.~M.; Safieva,~R.~Z.; Lomakina,~E.~I. Gasoline classification using
  near infrared (NIR) spectroscopy data: Comparison of multivariate techniques.
  \emph{Analytica Chimica Acta} \textbf{2010}, \emph{671}, 27--35\relax
\mciteBstWouldAddEndPuncttrue
\mciteSetBstMidEndSepPunct{\mcitedefaultmidpunct}
{\mcitedefaultendpunct}{\mcitedefaultseppunct}\relax
\EndOfBibitem
\bibitem[Misaki \latin{et~al.}(2010)Misaki, Kim, Bandettini, and
  Kriegeskorte]{misaki2010comparison}
Misaki,~M.; Kim,~Y.; Bandettini,~P.~A.; Kriegeskorte,~N. Comparison of
  multivariate classifiers and response normalizations for pattern-information
  fMRI. \emph{Neuroimage} \textbf{2010}, \emph{53}, 103--118\relax
\mciteBstWouldAddEndPuncttrue
\mciteSetBstMidEndSepPunct{\mcitedefaultmidpunct}
{\mcitedefaultendpunct}{\mcitedefaultseppunct}\relax
\EndOfBibitem
\bibitem[Tran \latin{et~al.}(2006)Tran, Wehrens, and Buydens]{tran2006knn}
Tran,~T.~N.; Wehrens,~R.; Buydens,~L.~M. KNN-kernel density-based clustering
  for high-dimensional multivariate data. \emph{Computational Statistics \&
  Data Analysis} \textbf{2006}, \emph{51}, 513--525\relax
\mciteBstWouldAddEndPuncttrue
\mciteSetBstMidEndSepPunct{\mcitedefaultmidpunct}
{\mcitedefaultendpunct}{\mcitedefaultseppunct}\relax
\EndOfBibitem
\bibitem[Joachims(2005)]{joachims2005support}
Joachims,~T. A support vector method for multivariate performance measures.
  Proceedings of the 22nd international conference on Machine learning. 2005;
  pp 377--384\relax
\mciteBstWouldAddEndPuncttrue
\mciteSetBstMidEndSepPunct{\mcitedefaultmidpunct}
{\mcitedefaultendpunct}{\mcitedefaultseppunct}\relax
\EndOfBibitem
\bibitem[Hastie \latin{et~al.}(1995)Hastie, Buja, and
  Tibshirani]{hastie1995penalized}
Hastie,~T.; Buja,~A.; Tibshirani,~R. Penalized discriminant analysis. \emph{The
  Annals of Statistics} \textbf{1995}, 73--102\relax
\mciteBstWouldAddEndPuncttrue
\mciteSetBstMidEndSepPunct{\mcitedefaultmidpunct}
{\mcitedefaultendpunct}{\mcitedefaultseppunct}\relax
\EndOfBibitem
\bibitem[Clemmensen \latin{et~al.}(2011)Clemmensen, Hastie, Witten, and
  Ersb{\o}ll]{clemmensen2011sparse}
Clemmensen,~L.; Hastie,~T.; Witten,~D.; Ersb{\o}ll,~B. Sparse discriminant
  analysis. \emph{Technometrics} \textbf{2011}, \emph{53}, 406--413\relax
\mciteBstWouldAddEndPuncttrue
\mciteSetBstMidEndSepPunct{\mcitedefaultmidpunct}
{\mcitedefaultendpunct}{\mcitedefaultseppunct}\relax
\EndOfBibitem
\bibitem[Thomaz \latin{et~al.}(2006)Thomaz, Kitani, and
  Gillies]{thomaz2006maximum}
Thomaz,~C.~E.; Kitani,~E.~C.; Gillies,~D.~F. A Maximum Uncertainty LDA-based
  approach for Limited Sample Size problems-with application to Face
  Recognition. \emph{Journal of the Brazilian Computer Society} \textbf{2006},
  \emph{12}, 7--18\relax
\mciteBstWouldAddEndPuncttrue
\mciteSetBstMidEndSepPunct{\mcitedefaultmidpunct}
{\mcitedefaultendpunct}{\mcitedefaultseppunct}\relax
\EndOfBibitem
\bibitem[Fisher and Sun(2011)Fisher, and Sun]{fisher2011improved}
Fisher,~T.~J.; Sun,~X. Improved Stein-type shrinkage estimators for the
  high-dimensional multivariate normal covariance matrix. \emph{Computational
  Statistics \& Data Analysis} \textbf{2011}, \emph{55}, 1909--1918\relax
\mciteBstWouldAddEndPuncttrue
\mciteSetBstMidEndSepPunct{\mcitedefaultmidpunct}
{\mcitedefaultendpunct}{\mcitedefaultseppunct}\relax
\EndOfBibitem
\bibitem[Dudoit \latin{et~al.}(2002)Dudoit, Fridlyand, and
  Speed]{dudoit2002comparison}
Dudoit,~S.; Fridlyand,~J.; Speed,~T.~P. Comparison of discrimination methods
  for the classification of tumors using gene expression data. \emph{Journal of
  the American statistical association} \textbf{2002}, \emph{97}, 77--87\relax
\mciteBstWouldAddEndPuncttrue
\mciteSetBstMidEndSepPunct{\mcitedefaultmidpunct}
{\mcitedefaultendpunct}{\mcitedefaultseppunct}\relax
\EndOfBibitem
\bibitem[Guo \latin{et~al.}(2006)Guo, Hastie, and
  Tibshirani]{guo2006regularized}
Guo,~Y.; Hastie,~T.; Tibshirani,~R. Regularized linear discriminant analysis
  and its application in microarrays. \emph{Biostatistics} \textbf{2006},
  \emph{8}, 86--100\relax
\mciteBstWouldAddEndPuncttrue
\mciteSetBstMidEndSepPunct{\mcitedefaultmidpunct}
{\mcitedefaultendpunct}{\mcitedefaultseppunct}\relax
\EndOfBibitem
\bibitem[Kemsley(1996)]{kemsley1996discriminant}
Kemsley,~E. Discriminant analysis of high-dimensional data: a comparison of
  principal components analysis and partial least squares data reduction
  methods. \emph{Chemometrics and intelligent laboratory systems}
  \textbf{1996}, \emph{33}, 47--61\relax
\mciteBstWouldAddEndPuncttrue
\mciteSetBstMidEndSepPunct{\mcitedefaultmidpunct}
{\mcitedefaultendpunct}{\mcitedefaultseppunct}\relax
\EndOfBibitem
\bibitem[Brereton and Lloyd(2014)Brereton, and Lloyd]{brereton2014partial}
Brereton,~R.~G.; Lloyd,~G.~R. Partial least squares discriminant analysis:
  taking the magic away. \emph{Journal of Chemometrics} \textbf{2014},
  \emph{28}, 213--225\relax
\mciteBstWouldAddEndPuncttrue
\mciteSetBstMidEndSepPunct{\mcitedefaultmidpunct}
{\mcitedefaultendpunct}{\mcitedefaultseppunct}\relax
\EndOfBibitem
\bibitem[Wehrens and Mevik(2007)Wehrens, and Mevik]{wehrens2007pls}
Wehrens,~R.; Mevik,~B.-H. The pls package: principal component and partial
  least squares regression in R. \emph{Journal of Statistical Software}
  \textbf{2007}, \emph{18}\relax
\mciteBstWouldAddEndPuncttrue
\mciteSetBstMidEndSepPunct{\mcitedefaultmidpunct}
{\mcitedefaultendpunct}{\mcitedefaultseppunct}\relax
\EndOfBibitem
\bibitem[Zhang \latin{et~al.}(2006)Zhang, Berg, Maire, and Malik]{zhang2006svm}
Zhang,~H.; Berg,~A.~C.; Maire,~M.; Malik,~J. SVM-KNN: Discriminative nearest
  neighbor classification for visual category recognition. Computer Vision and
  Pattern Recognition, 2006 IEEE Computer Society Conference on. 2006; pp
  2126--2136\relax
\mciteBstWouldAddEndPuncttrue
\mciteSetBstMidEndSepPunct{\mcitedefaultmidpunct}
{\mcitedefaultendpunct}{\mcitedefaultseppunct}\relax
\EndOfBibitem
\bibitem[Fix and Hodges(1989)Fix, and Hodges]{fix1989discriminatory}
Fix,~E.; Hodges,~J.~L. Discriminatory analysis. Nonparametric discrimination:
  consistency properties. \emph{International Statistical Review/Revue
  Internationale de Statistique} \textbf{1989}, \emph{57}, 238--247\relax
\mciteBstWouldAddEndPuncttrue
\mciteSetBstMidEndSepPunct{\mcitedefaultmidpunct}
{\mcitedefaultendpunct}{\mcitedefaultseppunct}\relax
\EndOfBibitem
\bibitem[Suykens and Vandewalle(1999)Suykens, and Vandewalle]{suykens1999least}
Suykens,~J.~A.; Vandewalle,~J. Least squares support vector machine
  classifiers. \emph{Neural processing letters} \textbf{1999}, \emph{9},
  293--300\relax
\mciteBstWouldAddEndPuncttrue
\mciteSetBstMidEndSepPunct{\mcitedefaultmidpunct}
{\mcitedefaultendpunct}{\mcitedefaultseppunct}\relax
\EndOfBibitem
\bibitem[Cortes and Vapnik(1995)Cortes, and Vapnik]{cortes1995machine}
Cortes,~C.; Vapnik,~V. Machine learning. \emph{Support vector networks}
  \textbf{1995}, \emph{20}, 273--297\relax
\mciteBstWouldAddEndPuncttrue
\mciteSetBstMidEndSepPunct{\mcitedefaultmidpunct}
{\mcitedefaultendpunct}{\mcitedefaultseppunct}\relax
\EndOfBibitem
\bibitem[Wang and Xue(2014)Wang, and Xue]{wang2014multi}
Wang,~Z.; Xue,~X. Multi-class support vector machine. In \emph{Support Vector
  Machines Applications}; Springer, 2014; pp 23--48\relax
\mciteBstWouldAddEndPuncttrue
\mciteSetBstMidEndSepPunct{\mcitedefaultmidpunct}
{\mcitedefaultendpunct}{\mcitedefaultseppunct}\relax
\EndOfBibitem
\bibitem[Weston and Watkins(1998)Weston, and Watkins]{weston1998multi}
Weston,~J.; Watkins,~C. \emph{Multi-class support vector machines}; 1998\relax
\mciteBstWouldAddEndPuncttrue
\mciteSetBstMidEndSepPunct{\mcitedefaultmidpunct}
{\mcitedefaultendpunct}{\mcitedefaultseppunct}\relax
\EndOfBibitem
\bibitem[Wichern and Johnson(1992)Wichern, and Johnson]{wichern1992applied}
Wichern,~D.~W.; Johnson,~R.~A. \emph{Applied multivariate statistical
  analysis}; Prentice Hall New Jersey, 1992; Vol.~4\relax
\mciteBstWouldAddEndPuncttrue
\mciteSetBstMidEndSepPunct{\mcitedefaultmidpunct}
{\mcitedefaultendpunct}{\mcitedefaultseppunct}\relax
\EndOfBibitem
\bibitem[Devijver and Kittler(1982)Devijver, and Kittler]{devijver1982pattern}
Devijver,~P.~A.; Kittler,~J. \emph{Pattern recognition: A statistical
  approach}; Prentice hall, 1982\relax
\mciteBstWouldAddEndPuncttrue
\mciteSetBstMidEndSepPunct{\mcitedefaultmidpunct}
{\mcitedefaultendpunct}{\mcitedefaultseppunct}\relax
\EndOfBibitem
\bibitem[Rao(1948)]{rao1948utilization}
Rao,~C.~R. The utilization of multiple measurements in problems of biological
  classification. \emph{Journal of the Royal Statistical Society. Series B
  (Methodological)} \textbf{1948}, \emph{10}, 159--203\relax
\mciteBstWouldAddEndPuncttrue
\mciteSetBstMidEndSepPunct{\mcitedefaultmidpunct}
{\mcitedefaultendpunct}{\mcitedefaultseppunct}\relax
\EndOfBibitem
\bibitem[Yu and Yang(2001)Yu, and Yang]{yu2001direct}
Yu,~H.; Yang,~J. A direct LDA algorithm for high-dimensional data—with
  application to face recognition. \emph{Pattern recognition} \textbf{2001},
  \emph{34}, 2067--2070\relax
\mciteBstWouldAddEndPuncttrue
\mciteSetBstMidEndSepPunct{\mcitedefaultmidpunct}
{\mcitedefaultendpunct}{\mcitedefaultseppunct}\relax
\EndOfBibitem
\bibitem[Campbell(1980)]{campbell1980shrunken}
Campbell,~N.~A. Shrunken estimators in discriminant and canonical variate
  analysis. \emph{Applied Statistics} \textbf{1980}, 5--24\relax
\mciteBstWouldAddEndPuncttrue
\mciteSetBstMidEndSepPunct{\mcitedefaultmidpunct}
{\mcitedefaultendpunct}{\mcitedefaultseppunct}\relax
\EndOfBibitem
\bibitem[Peck and Van~Ness(1982)Peck, and Van~Ness]{peck1982use}
Peck,~R.; Van~Ness,~J. The use of shrinkage estimators in linear discriminant
  analysis. \emph{IEEE Transactions on Pattern Analysis and Machine
  Intelligence} \textbf{1982}, 530--537\relax
\mciteBstWouldAddEndPuncttrue
\mciteSetBstMidEndSepPunct{\mcitedefaultmidpunct}
{\mcitedefaultendpunct}{\mcitedefaultseppunct}\relax
\EndOfBibitem
\bibitem[Cawley and Talbot(2003)Cawley, and Talbot]{cawley2003efficient}
Cawley,~G.~C.; Talbot,~N.~L. Efficient leave-one-out cross-validation of kernel
  fisher discriminant classifiers. \emph{Pattern Recognition} \textbf{2003},
  \emph{36}, 2585--2592\relax
\mciteBstWouldAddEndPuncttrue
\mciteSetBstMidEndSepPunct{\mcitedefaultmidpunct}
{\mcitedefaultendpunct}{\mcitedefaultseppunct}\relax
\EndOfBibitem
\bibitem[P{\'e}rez-Enciso and Tenenhaus(2003)P{\'e}rez-Enciso, and
  Tenenhaus]{perez2003prediction}
P{\'e}rez-Enciso,~M.; Tenenhaus,~M. Prediction of clinical outcome with
  microarray data: a partial least squares discriminant analysis (PLS-DA)
  approach. \emph{Human genetics} \textbf{2003}, \emph{112}, 581--592\relax
\mciteBstWouldAddEndPuncttrue
\mciteSetBstMidEndSepPunct{\mcitedefaultmidpunct}
{\mcitedefaultendpunct}{\mcitedefaultseppunct}\relax
\EndOfBibitem
\bibitem[Martens(2001)]{martens2001reliable}
Martens,~H. Reliable and relevant modelling of real world data: a personal
  account of the development of PLS regression. \emph{Chemometrics and
  intelligent laboratory systems} \textbf{2001}, \emph{58}, 85--95\relax
\mciteBstWouldAddEndPuncttrue
\mciteSetBstMidEndSepPunct{\mcitedefaultmidpunct}
{\mcitedefaultendpunct}{\mcitedefaultseppunct}\relax
\EndOfBibitem
\bibitem[Wold(2001)]{wold2001personal}
Wold,~S. Personal memories of the early PLS development. \emph{Chemometrics and
  Intelligent Laboratory Systems} \textbf{2001}, \emph{58}, 83--84\relax
\mciteBstWouldAddEndPuncttrue
\mciteSetBstMidEndSepPunct{\mcitedefaultmidpunct}
{\mcitedefaultendpunct}{\mcitedefaultseppunct}\relax
\EndOfBibitem
\bibitem[Davies and Bouldin(1979)Davies, and Bouldin]{davies1979cluster}
Davies,~D.~L.; Bouldin,~D.~W. A cluster separation measure. \emph{IEEE
  transactions on pattern analysis and machine intelligence} \textbf{1979},
  224--227\relax
\mciteBstWouldAddEndPuncttrue
\mciteSetBstMidEndSepPunct{\mcitedefaultmidpunct}
{\mcitedefaultendpunct}{\mcitedefaultseppunct}\relax
\EndOfBibitem
\bibitem[Izenman(2013)]{izenman2013linear}
Izenman,~A.~J. Linear discriminant analysis. In \emph{Modern multivariate
  statistical techniques}; Springer, 2013; pp 237--280\relax
\mciteBstWouldAddEndPuncttrue
\mciteSetBstMidEndSepPunct{\mcitedefaultmidpunct}
{\mcitedefaultendpunct}{\mcitedefaultseppunct}\relax
\EndOfBibitem
\bibitem[Guyon \latin{et~al.}(1998)Guyon, Makhoul, Schwartz, and
  Vapnik]{guyon1998size}
Guyon,~I.; Makhoul,~J.; Schwartz,~R.; Vapnik,~V. What size test set gives good
  error rate estimates? \emph{IEEE Transactions on Pattern Analysis and Machine
  Intelligence} \textbf{1998}, \emph{20}, 52--64\relax
\mciteBstWouldAddEndPuncttrue
\mciteSetBstMidEndSepPunct{\mcitedefaultmidpunct}
{\mcitedefaultendpunct}{\mcitedefaultseppunct}\relax
\EndOfBibitem
\bibitem[Rao and Scott(1981)Rao, and Scott]{rao1981analysis}
Rao,~J.~N.; Scott,~A.~J. The analysis of categorical data from complex sample
  surveys: chi-squared tests for goodness of fit and independence in two-way
  tables. \emph{Journal of the American statistical association} \textbf{1981},
  \emph{76}, 221--230\relax
\mciteBstWouldAddEndPuncttrue
\mciteSetBstMidEndSepPunct{\mcitedefaultmidpunct}
{\mcitedefaultendpunct}{\mcitedefaultseppunct}\relax
\EndOfBibitem
\bibitem[Ripley(2007)]{knn_package}
Ripley,~B.~D. \emph{Pattern recognition and neural networks}; Cambridge
  university press, 2007\relax
\mciteBstWouldAddEndPuncttrue
\mciteSetBstMidEndSepPunct{\mcitedefaultmidpunct}
{\mcitedefaultendpunct}{\mcitedefaultseppunct}\relax
\EndOfBibitem
\bibitem[Fan \latin{et~al.}(2005)Fan, Chen, and Lin]{svm_package}
Fan,~R.-E.; Chen,~P.-H.; Lin,~C.-J. Working set selection using second order
  information for training support vector machines. \emph{Journal of machine
  learning research} \textbf{2005}, \emph{6}, 1889--1918\relax
\mciteBstWouldAddEndPuncttrue
\mciteSetBstMidEndSepPunct{\mcitedefaultmidpunct}
{\mcitedefaultendpunct}{\mcitedefaultseppunct}\relax
\EndOfBibitem
\bibitem[Silva \latin{et~al.}(2015)Silva, Silva, and
  Suggests]{LDA_sparse_package}
Silva,~A. P.~D.; Silva,~M. A. P.~D.; Suggests,~M. Package ‘HiDimDA’.
  \textbf{2015}, \relax
\mciteBstWouldAddEndPunctfalse
\mciteSetBstMidEndSepPunct{\mcitedefaultmidpunct}
{}{\mcitedefaultseppunct}\relax
\EndOfBibitem
\bibitem[Hubert and Arabie(1985)Hubert, and Arabie]{hubert1985comparing}
Hubert,~L.; Arabie,~P. Comparing partitions. \emph{Journal of classification}
  \textbf{1985}, \emph{2}, 193--218\relax
\mciteBstWouldAddEndPuncttrue
\mciteSetBstMidEndSepPunct{\mcitedefaultmidpunct}
{\mcitedefaultendpunct}{\mcitedefaultseppunct}\relax
\EndOfBibitem
\end{mcitethebibliography}
\end{document}